\documentclass[11pt,draftcls,onecolumn,journal]{IEEEtran}
\usepackage{amsmath,amsfonts}
\usepackage{algorithm}
\usepackage{array}
\usepackage[caption=false,font=normalsize,labelfont=sf,textfont=sf]{subfig}
\usepackage{textcomp}
\usepackage{stfloats}
\usepackage{url}
\usepackage{verbatim}
\usepackage{graphicx}
\usepackage{cite}
\usepackage{hyperref}
\usepackage{booktabs}
\usepackage{framed}
\usepackage{tcolorbox}
\usepackage{footnotehyper}
\usepackage{algpseudocode}
\makesavenoteenv{tcolorbox} 
\usepackage{amssymb,amsmath}
\usepackage{graphics}
\usepackage{float}
\usepackage{color}
\usepackage{dsfont}
\usepackage{bm}
\usepackage{upgreek}
\usepackage{arydshln}
\usepackage{enumerate}
\usepackage{amsthm}
\usepackage{graphicx}
\usepackage{threeparttable}
\usepackage{caption}
\usepackage{multirow}
\usepackage{color}
\usepackage{xfrac}
\usepackage{array}
\usepackage{nomencl}
\usepackage{hyphenat}
\usepackage{url}
\usepackage[font=small]{caption}
\usepackage{hyphenat}
\usepackage{setspace}
\usepackage{cancel}
\usepackage[dvipsnames]{xcolor}

\newcommand{\bs}[1]{\boldsymbol{#1}}

\newcommand{\virg}[1]{\textquotedblleft#1\textquotedblright}

\newcommand\e[1]{E\{ #1 \} }
\newcommand\eg[1]{E\left\lbrace  #1 \right\rbrace  }
\newcommand\evg[1]{{E_0\left\lbrace  #1 \right\rbrace  }}
\newcommand\ev[1]{{E_0\{ #1 \} }}
\newcommand{\tonde}[1]{\left( #1 \right)  }

\newcommand{\graffe}[1]{\left\lbrace   #1 \right\rbrace   }


\newcommand{\RV}[1]{{\color{black}#1}}

\begin{document}
%
\title{The Symmetric Location Problem: a Song of Efficiency and Robustness}
%
\author{Stefano~Fortunati

\thanks{Stefano~Fortunati is with SAMOVAR, Télécom SudParis, Institut Polytechnique de Paris, Palaiseau, France.}
}
%
\markboth{IEEE Signal Processing Magazine,~Vol.~XX, No.~XX, June~2024}%
{Fortunati: Author Guidelines for Columns \& Forum Articles of IEEE SPM}


\maketitle

%
%

\section*{Scope}
The aim of this Lecture Note is to introduce the Signal Processing (SP) community to a powerful yet still under-utilised tool: the semiparametric statistics. In short, the semiparametric framework allows us to estimate or perform hypothesis testing on a \textit{finite-dimensional parameter} $\bs{\theta} \in \Theta \subseteq \mathbb{R}^p$ in the presence of an \textit{infinite-dimensional nuisance parameter} $g \in \mathcal{S}$ (i.e. a function), such as the density of the noise. Clearly, this framework is general enough to include almost every SP application. 

Remarkably, as the title suggests drawing on George R.\ R.\ Martin’s famous book series, the greatest advantage of semiparametric statistics over parametric and non-parametric ones lies in the fact that it is able to reconcile two seemingly dichotomous concepts: \textit{statistical efficiency} and \textit{distributional robustness}. 

To explain exactly what this means, in this Lecture Note we will focus our attention on the famous and fundamental symmetric location problem.

\section*{Relevance}
The problem of estimating the center of symmetry, i.e. the \textit{location}, of an univariate simmetric probability density function (pdf) is one of the oldest problems addressed in mathematical statistics. Given its ubiquity in so many applications, including those in SP, this problem has served over the years as a test bed for various statistical theories, ranging from parametric, robust, and non-parametric statistics. In the same vein, in this Lecture Note we shall test the tools of semiparametric statistics on it. Not only is this of interest in its own right, but it will also give us the opportunity to demonstrate the semiparametric machinery in action and the great potential it offers. To this end, let us provide some examples of semiparametric estimation problems relevant in SP applications to further emphasize the importance that this approach may have within the SP community \cite{Chap_sem}:
	\begin{itemize}
		\item Any linear and non-linear regression problem can be considered as a semiparametric estimation problem where the infinite-dimensional nuisance parameter $g \in \mathcal{S}$ is the noise density,
		\item In inference procedures involving missing data, the nuisance function $g \in \mathcal{S}$ may characterize the density of the \textit{unobservable} data,
		\item In time series analysis, $g \in \mathcal{S}$ may represent the density of the innovation process underlying the data generation.
		\item In heavy-tailed data processing, $g \in \mathcal{S}$ may be used as a more realistic substitute for the bell-shaped Gaussian distribution. 
	\end{itemize} 
Since $g \in \mathcal{S}$ is left unspecified, a semiparametric model generally contains less information than a classical parametric model. This will generally result in a loss of statistical efficiency in the estimation of the parameter vector of interest $\bs{\theta} \in \Theta \subseteq \mathbb{R}^p$ compared with the classical, but often unrealistic, parametric approach. A SP practitioner should, in fact, be aware that this loss of efficiency is merely apparent, as a parametric model is often too rigid and limited to describe reality. Consequently, the more conservative semiparametric approach should always be preferred to avoid the risk of running straight into a problem of model misspecification, which can render a parametric estimation algorithm completely unreliable.

Remarkably, as we shall see in the case of the symmetric location problem, there are examples where leaving the nuisance function $g \in \mathcal{S}$ unspecified does not result in any loss of efficiency with respect to the parametric case in which $g$ is assumed to be a priori known! This is an ideal situation in which we can have the (distributional) robustness of a semiparametric procedure and the statistical efficiency of a parametric approach: the proverbial situation in which you can have your cake and eat it too.\footnote{Others \virg{best of both worlds} semiparametric cases are discussed in \cite{For_Delmas_Ollila_TIT} in the context of elliptical data.}

Let us now consider the property that we expect from a semiparametric estimator. Since the nuisance function $g \in \mathcal{S}$ is left unspecified, we would like that such estimator (more precisely, its pdf) is completely invariant with respect to $g \in \mathcal{S}$. In other words, we would like to use the estimator \virg{blindly}: whatever $g \in \mathcal{S}$ characterizes the data generating process, we could rely on the semiparametric estimator because its performance will not depend on $g \in \mathcal{S}$. This is exactly the \textit{distribution-freeness} (or \textit{distributional robustness}) property.

The semiparametric approach to the symmetric location problem offers us a perfect means to clarify all the points just made. Moreover, from a practical point of view, the symmetric location problem can be found (in various forms) in countless areas of SP: source localization (GPS/GNSS positioning or locating interference/jamming sources), time synchronization (distributed radar, sonar or microphones, wireless sensor networks and power-grid monitoring), array processing (waveform synthesis and direction of arrival estimation), just to name a few. Furthermore, it is important to note that the methodology we will develop for this specific problem can be extended to much more general semiparametric estimation problems, such as the estimation of the location vector and covariance matrix in elliptical data \cite{For_Delmas_Ollila_TIT}. 

\section*{Prerequisites}
The prerequisites for a thorough understanding of the semiparametric estimation theory are substantial, and we are well aware that they generally do not fall within the traditional background of the SP community. However, our aim here is not to provide a tutorial at the end of which the reader will be able to solve a semiparametric estimation problem from scratch. Instead, our aim is to highlight that statistics has evolved well beyond the foundational SP textbooks (such as those by van Trees and Kay) revealing a rich landscape of new ideas and techniques awaiting to be explored.

\section*{Problem statement}
Let $X_1, \dots, X_n$ be $n$ \textit{independent and identically distributed} (i.i.d.)\ observations such that: \footnote{\textit{Background}: Let $(\mathcal{X},\mathfrak{B}(\mathcal{X}),P_0)$ be a probability space where the sample space $\mathcal{X}$ is (a subset of) $\mathbb{R}$, $\mathfrak{B}(\mathcal{X})$ is the Borel $\sigma$-algebra on $\mathcal{X}$ and $P_0$ is a probability measure. Moreover, $P_0$ is assumed to be \textit{absolutely continuous} with respect to (w.r.t.)\ the Lebesgue measure $\lambda$ on $\mathbb{R}$ with density $p_0$ such that $dP_0=p_0d\lambda$. Let $f:\mathcal{X}\rightarrow \mathbb{R}$ be a $\mathfrak{B}(\mathcal{X})$-measurable function, then $E_0\{f\} \triangleq \int fdP_0$ indicates its expectation w.r.t.\ $P_0$. To avoid any confusion, in the rest of this note we indicate as $X_i$ the random variable corresponding to an observation, while $x_i$ denotes its realization or the argument of a relevant pdf. Let $X_l$ be a sequence of random variables, then we write \textit{i)} $X_l = o_{L_2}(1)$ if $\lim_{l\rightarrow \infty}E\{|X_l|^2\}=0$ (\textit{mean-square convergence}), \textit{ii)} $X_l = o_{P_0}(1)$ if $\lim_{l\rightarrow \infty}\mathrm{Pr}\graffe{|X_l|\geq\epsilon}=0,\forall \epsilon>0$ (\textit{convergence in $P_0$-probability to 0}), \textit{iii)} $X_l = O_{P_0}(1)$ if for any $\epsilon > 0$, there exists a finite $M>0$ and a finite $L>0$, s.t. $\mathrm{Pr}\graffe{|X_l|>M}<\epsilon,\forall l>L$ (\textit{stochastic boundedness}) and \textit{iv)} $X_l \overset{d}{\rightarrow} X$ means that the sequence $X_l$ \textit{converges in distribution} to the random variable $X$.}
\begin{equation}\label{sim_p_0}
	\mathbb{R} \ni X_i \sim p_0(x) = g_0(x- \theta_0),
\end{equation}
where $g_0$ belongs to the set of \textit{non-vanishing} (i.e., always greater than 0), \textit{symmetric} pdfs:
\begin{equation}
	\mathcal{S} \triangleq \graffe{g:\mathbb{R}\rightarrow\mathbb{R}^+| g(x) = g(-x) },
\end{equation} 
with cumulative distribution function (cdf) given by $G(x) = \int_{-\infty}^{x}g(\alpha)d\alpha$, for all $g \in \mathcal{S}$.

We aim to estimate the center of symmetry $\theta_0$ in a \emph{semiparametric sense}, that is without specifying any $g \in \mathcal{S}$ as generating density of the data. In particular, we will consider $g \in \mathcal{S}$ as an infinite-dimensional (functional) nuisance parameter. Then, the relevant semiparametric model can be cast as follows:
\begin{equation}\label{sem_par}
	\mathcal{P}_{\theta,g} \triangleq \graffe{p_{\theta,g}(x)= g(x- \theta)| \theta \in \mathbb{R}, g \in  \mathcal{S}}.
\end{equation}
Note that the classical parametric location problem is a particular case of the semiparametric one, since it is based on the parametric model:
\begin{equation}\label{par_mod}
	\mathcal{P}_{\theta} \triangleq \graffe{p_{\theta}(x)= g_0(x- \theta)|\theta \in \mathbb{R}} \subset \mathcal{P}_{\theta,g}.
\end{equation}
For further reference, we state that $\mathcal{P}_{\theta}$ is a \textit{parametric sub-model} of $\mathcal{P}_{\theta,g}$. We also introduce the \textit{nonparametric sub-model} $\mathcal{P}_{g}$ as:
\begin{equation}\label{nonpar_mod}
	\mathcal{P}_{g} \triangleq \graffe{p(x)= g(x- \theta_0)|g \in \mathcal{S}} \subset \mathcal{P}_{\theta,g}.
\end{equation}
We will show two surprising properties of the semiparametric symmetric location model. 

Let $\hat{\theta}$ be an estimator of $\theta_0$, then: 
\begin{itemize}
	\item The semiparametric lower bound to the Mean Squared Error (MSE) $\ev{(\hat{\theta}-\theta_0)^2}$ of any consistent estimator $\hat{\theta}$ of $\theta_0$ derived in the semiparametric model $\mathcal{P}_{\theta,g}$ is equal to the classical parametric bound derived in $\mathcal{P}_{\theta}$. In other words, knowing or not knowing the true density $g_0$ does not have any impact on the (asymptotic) efficiency bound. This result is proven in the Section \virg{From $\mathcal{P}_{\theta}$ to $\mathcal{P}_{\theta,g}$: adaptivity and the semiparametric efficiency bound}.
	\item It is possible to derive a \textit{$g_0$-free} and (almost) \textit{distribution-free}, or \textit{distributionally robust}, estimator $\hat{\theta}$ that is asymptotically Gaussian-distributed with an (asymptotic) error covariance that can be explicitly evaluated and compared to the (semi)parametric lower bound in order to assess its efficiency for any possible $g \in \mathcal{S}$. These findings will be presented in the Section \virg{One-step (OS) $R$-estimator}.
\end{itemize}

It is worth emphasising that, even though the two properties mentioned above hold for the symmetric location problem as we shall show in the next section, this does not mean that they hold for any semiparametric model. 

Before we continue, it is worth clarifying the terminology we are going to use:
	\begin{itemize}
		\item A \textit{$g_0$-free} estimator is an estimator that can be constructed without prior knowledge of $g_0$. \RV{Clearly, this does not imply identical estimation accuracy under all nuisance functions $g \in \mathcal{S}$. For example, the sample mean is a $g_0$-free estimator of the location $\theta_0$ but its performance its performance deteriorates significantly when the data-generating density deviates from the Gaussian distribution.}
		\item The \textit{distribution-freeness}, or \textit{distributional robustness}, dictates that the density of the estimator is invariant to $g_0$. \RV{This is not the case, for example, for the sample mean.}
		\item The \textit{semiparametric efficiency} implies that the estimator achieves the minimal possible asymptotic error variance across the entire family of densities $\mathcal{S}$.
		\item The \textit{robustness to outliers} characterizes the resilience to the data contamination. This last type of robustness will not be covered in this Lecture Note.
	\end{itemize}

\section*{Solution}
\subsection*{Local Asymptotic Normality of the parametric sub-model $\mathcal{P}_{\theta}$}

A milestone in modern statistics is the concept of Local Asymptotic Normality (LAN), introduced by Lucien Le Cam \cite{LeCam60} (see also \cite[Ch.7]{vaart_1998}). The LAN property offers a unifying framework that revisits familiar concepts, such as the \textit{score} and \textit{Fisher Information}, in a far more general setting. To balance rigor with clarity, in this Lecture Note we decided to adopt over-restrictive regularity conditions, grounding the discussion in familiar, well-understood quantities.

Let us start from the parametric sub-model $\mathcal{P}_{\theta}$ in \eqref{par_mod}, where the generating density $g_0 \in \mathcal{S}$ is assumed to be known. Then, we can introduce the well-known parametric \textit{score} for the center of symmetry $\theta_0$ as:
\begin{equation}\label{s_t}
	s_{\theta_0}(x) \triangleq \left. \frac{d }{d \theta}\ln g_0(x- \theta)\right|_{\theta = \theta_0} = \varphi_{g_0}(x- \theta_0),
\end{equation}
where we defined the function:
\begin{equation}\label{phi_fun}
	\varphi_g(u) \triangleq - \frac{g'(u)}{g(u)}, \quad g \in \mathcal{S}.
\end{equation}
Note that, since $g$ is even, $g'$ is odd, that is $g'(u) = -g'(-u)$ and consequently $\varphi_g(u)$ is an odd function for all $g \in \mathcal{S}$. This immediately implies that the score $s_{\theta_0}$ is an odd function of $(x-\theta_0)$. We can also introduce the parametric \textit{Fisher Information} (FI) for the center of symmetry $\theta_0$ as:
\begin{equation}\label{FI}
	I(\theta_0) \triangleq \evg{s_{\theta_0}^2(X)} = \evg{\varphi_{g_0}^2(X- \theta_0)} = - \evg{\varphi'_{g_0}(X- \theta_0)}.
\end{equation}

In the rest of this Lecture Note, we will always assume that \cite[Lemma 7.6]{vaart_1998}:
\begin{itemize}
	\item[A1)] The map $\theta \mapsto \sqrt{p_{\theta}(x)}$ is continuously differentiable for every $x \in \mathcal{X}$,
	\item[A2)] $I(\theta)$ is well-defined and continuous in $\theta \in \mathbb{R}$.
\end{itemize} 
Under A1 and A2, an important property of the score that will be used in the following is that:
\begin{equation}\label{sc_0}
	\ev{s_{\theta_0}(X)} = 0.
\end{equation}
Moreover, again under A1 and A2, we have that the sequence of parametric sub-models: 
\begin{equation}
	\mathcal{P}^n_{\theta} \triangleq \graffe{p_\theta(x_1,\ldots,x_n)=\prod\nolimits_{i=1}^{n}g(x_i- \theta)| \theta \in \mathbb{R}},
\end{equation}
is \textit{locally asymptotically normal} at a point $\theta_0$, that is \cite[Theo. 7.2]{vaart_1998}, for $h \in \mathbb{R}$, as $n \rightarrow \infty$:
\begin{equation}\label{LAN}
	\ln\prod\nolimits_{i=1}^{n}\frac{p_{\theta_0 + h/\sqrt{n}}}{p_{\theta_0}}(X_i) = \Delta_{g_0,n}(\theta_0)h - \frac{1}{2}I(\theta_0)h^2 + o_{P_0}(1),
\end{equation} 
where, in the Le Cam' terminology, the term $\Delta_{g_0,n}(\theta_0)$ is called \textit{central sequence} and it is given by:
\begin{equation}\label{cen_seq}
	\Delta_{g_0,n}(\theta_0) \triangleq \frac{1}{\sqrt{n}}\sum\nolimits_{i=1}^{n}s_{\theta_0}(X_i).
\end{equation}
A fundamental property of the central sequence that we will use throughout this Lecture Note is that $\Delta_{g_0,n}(\theta_0)$ is asymptotically normal with zero mean and variance $I(\theta_0)$:
\begin{equation}\label{cs_dist}
		\Delta_{g_0,n}(\theta_0) \overset{d}{\rightarrow} \mathcal{N}(0,I(\theta_0)).
\end{equation}

The intuition behind \eqref{LAN} and \eqref{cs_dist} is as follows. Let us introduce the log-likelihood function
\begin{equation}\label{log_lik}
	l(\theta) \triangleq \sum_{i=1}^{n} \ln p_\theta(x_i) = \sum_{i=1}^{n} \ln g_0(x_i- \theta).
\end{equation}	
Then the LHS of \eqref{LAN} can be expressed, for i.i.d. data and under appropriate regularity assumptions that are stronger that A1 and A2, by using the Taylor expansion of $l(\theta_0 + h/\sqrt{n})$ up to the second order:
\begin{equation}
	\begin{split}
		l(\theta_0 + h/\sqrt{n}) - l(\theta_0) &= \frac{h}{\sqrt{n}}l'(\theta_0) + \frac{h^2}{n}l''(\theta_0) + o_{P_0}(1)\\
		& =  \frac{h}{\sqrt{n}}\sum_{i=1}^{n}\varphi_{g_0}(X_i - \theta_0) +  \frac{h^2}{n}\sum_{i=1}^{n}\varphi'_{g_0}(X_i- \theta_0)+ o_{P_0}(1).
	\end{split}
\end{equation}
Given \eqref{FI}, from the Law of Large Number, we have that:
\begin{equation}
	\frac{1}{n}\sum_{i=1}^{n}\varphi'_{g_0}(X_i- \theta_0) = - I(\theta_0) + o_{P_0}(1),
\end{equation} 
and consequently, \eqref{LAN} follows. Moreover, \eqref{cs_dist} can be obtained from \eqref{s_t} and \eqref{sc_0} and from the direct application of the Central Limit Theorem:
\begin{equation}
	\frac{1}{\sqrt{n}}\sum_{i=1}^{n}\varphi_{g_0}(X_i - \theta_0) \overset{d}{\rightarrow} \mathcal{N}(0,I(\theta_0)).
\end{equation}
It is worth underling however that the relation \eqref{LAN} can be proved under much weaker regularity conditions then the ones required for the above-mentioned Taylor expansion and, more important, in the presence of dependent data. 

\subsection*{From $\mathcal{P}_{\theta}$ to $\mathcal{P}_{\theta,g}$: adaptivity and the semiparametric efficiency bound}
While our journey has been relatively smooth so far, the road begins to get steeper now that we must define the counterparts of the score and the FI for the semiparametric model $\mathcal{P}_{\theta,g}$ where $g \in \mathcal{S}$ is an infinite-dimensional unknown parameter. How do we handle functional parameters in this context? The bridge between classical parametric and semiparametric theory lies in the concept of Hilbert space, i.e.\ an inner product space that is also a complete metric space with respect to the distance function induced by the inner product. Since the tools associated with this space are dimension-agnostic, they apply equally to finite and infinite dimensional parameters. This means we can unify the treatment of both types of parameters within the same mathematical framework. Specifically, we need the following three ingredients:
\begin{itemize}
	\item The Hilbert space $\mathcal{H}$ of all zero-mean and finite variance, random functions $h$ of the observations:
	\begin{equation}\label{H_set}
		\mathcal{H} \triangleq \graffe{h:\mathbb{R}\rightarrow \mathbb{R}|\ev{h(X)} = 0, \ev{h^2(X)}< +\infty}.
	\end{equation}
	This is the \virg{ambient space} in which an estimation problem can be embedded.
	\item A notion of \textit{nuisance tangent space} $\mathcal{T} \subseteq \mathcal{H}$ for a parametric or semiparametric model. This subset of $\mathcal{H}$ characterizes the nuisance parameter in a geometrical way.
	\item An orthogonal projection operator of a generic function $h \in \mathcal{H}$ onto $\mathcal{T}$, i.e.\ $\Pi(h|\mathcal{T}) \in \mathcal{T}$.
\end{itemize}
For a formal definition of these concepts we refer the reader to the reference book \cite[Ch. 3]{BKRW} while a tutorial and accessible presentation can be found in \cite[Sect. 5.2]{Chap_sem}. Here we will limit ourselves to provide a simple and intuitive explanation with a direct link to the symmetric location problem. First of all, we can notice that, under A1 and A2, the score $s_{\theta_0}$ in \eqref{s_t} belongs to $\mathcal{H}$.\footnote{In fact $\ev{s_{\theta_0}(X)} = 0$ and $\ev{s^2_{\theta_0}(X)} = I(\theta_0) < \infty$.} From the way it is defined, it is clear that this parametric score characterizes the problem of estimating $\theta_0$ when $g_0$ is known. However, since in our semiparametric problem $g_0$ is not known a priori, we must subtract from $s_{\theta_0}$ in \eqref{s_t} the amount of missing information associated with $g_0$. 

This is when the magic of the geometric approach comes into play. In fact, we can introduce the \textit{semiparametric efficient score} $\bar{s}_{\theta_0}$ as the residual of the parametric score $s_{\theta_0}$ after projecting it onto the nuisance tangent space $\mathcal{T}_{g_0}$ associated to the nuisance function $g_0 \in \mathcal{S}$:
\begin{equation}\label{def_eff_s_t}
	\boxed{\bar{s}_{\theta_0} \triangleq  s_{\theta_0} - \Pi(s_{\theta_0}|\mathcal{T}_{g_0})}.
\end{equation}
We can now define the \textit{semiparametric efficient FI} (EFI) as:
\begin{equation}\label{def_eff_FI}
	\boxed{\bar{I}(\theta_0|g_0) \triangleq \ev{\bar{s}^2_{\theta_0}(X)}}.
\end{equation}
The EFI represents the amount of information that the observations provide on the parameter of interest $\theta_0$ after accounting for the lack of information due to the fact that we do not know the true density $g_0$. 

Intuitively, one would naturally expect the EFI $\bar{I}(\theta_0|g_0)$ to be smaller than the FI $I(\theta_0)$. Surprisingly, this is not always the case. Remarkably, for the symmetric location problem \footnote{It is worth to stress again that this is not a general property of any semiparametric model.}, we have that $\bar{I}(\theta_0|g_0)=I(\theta_0)$! In other words, not knowing the data-generating density $g_0$ does not result in any loss of the amount of information that the observations provide about the center of symmetry $\theta_0$. When this happens, the semiparametric estimation problem is said to be \textit{adaptive}. 

To illustrate the point just made, though without any claim of completeness, we will now provide an explicit expression of the nuisance tangent space $\mathcal{T}_{g_0}$ and its corresponding projection operator $\Pi(h|\mathcal{T}_{g_0})$. Readers interested in understanding all the mathematical details can find them in \cite[Sect.\ 3.2]{BKRW}. Following \cite[Sect.\ 3.2, Ex.\ 2]{BKRW}, we have that the tangent space associated to the nuisance density $g_0$ in the semiparametric model $\mathcal{P}_{\theta,g}$ in \eqref{sem_par} is:
\begin{equation}
	\mathcal{T}_{g_0}  = \graffe{h \in \mathcal{H}| h(x) = h(|x-\theta_0|)}.
\end{equation}
Moreover, from \cite[App.\ 3.1]{BKRW} we have that the relevant projection operator is given by:
\begin{equation}\label{proj}
	\Pi(h|\mathcal{T}_{g_0}) = \frac{1}{2}(h(x)+h(2\theta_0 -x)), \quad \forall h \in \mathcal{H}.
\end{equation}
We now have all the ingredients to evaluate the semiparametric efficient score in \eqref{def_eff_s_t}. In fact, from \eqref{s_t} and \eqref{proj}, we have that:
\begin{equation}\label{eff_s_t}
	\bar{s}_{\theta_0}(x) = s_{\theta_0}(x) - \frac{1}{2}(s_{\theta_0}(x)+s_{\theta_0}(2\theta_0 -x)) = s_{\theta_0}(x),
\end{equation}
where the last equality follows from the oddity w.r.t. $(x-\theta_0)$ of $s_{\theta_0}$. We have found a quite surprising result: \textit{the semiparametric efficient score is equal to the parametric score for the center of symmetry $\theta_0$}. As a direct consequence, we have that the semiparametric EFI is equal to the parametric FI:
\begin{equation}\label{eff_FI}
	\bar{I}(\theta_0|g_0) = I(\theta_0) = \evg{\varphi_{g_0}^2(X- \theta_0)}.
\end{equation}
We can then conclude that the semiparametric symmetric location model $\mathcal{P}_{\theta,g}$ in \eqref{sem_par} is \textit{adaptive} for $\theta_0$ in the sense that it contains the same amount of information of its parametric sub-model $\mathcal{P}_{\theta}$ in \eqref{par_mod}.  

A question naturally arises: in classical parametric statistics, we know that the inverse of the FI $I(\theta_0)$ provides the Cramér-Rao bound (CRB), which is a lower bound on the MSE of any (unbiased) estimator $\hat{\theta}$ of $\theta_0$ in $\mathcal{P}_{\theta}$. Can we define a similar lower bound as the inverse of the EFI $\bar{I}(\theta_0|g_0)$ for the semiparametric estimation problem in $\mathcal{P}_{\theta,g}$? The answer is yes, but the proof of this result is far from simple and is based on the semiparametric extension of the celebrated Hájek–Le Cam convolution theorem \cite[Sect.\ 3.3, Th.\ 2]{BKRW}, \cite[Th.\ 25.21]{vaart_1998}. In particular, it can be shown that any $\sqrt{n}$-\textit{consistent} semiparametric estimator $\hat{\theta}_n$ of $\theta_0$ derived in $\mathcal{P}_{\theta,g}$ satisfies the following inequality:
\begin{equation}\label{LB}
	\boxed{\lim_{n\rightarrow \infty} \evg{\big[\sqrt{n}(\hat{\theta}_n-\theta_0)\big]^2}\geq \bar{I}(\theta_0|g_0)^{-1}.}
\end{equation}
It is important to note here that, thanks to the adaptivity property of the symmetric location model, such semiparametric lower bound is equal to the parametric CRB, i.e. $\bar{I}(\theta_0|g_0)^{-1} = I(\theta_0)^{-1} = \mathrm{CRB}(\theta_0)$. 

And now the million-dollar question: can we derive a semiparametric estimator, that is a \textit{$g_0$- and distribution-free} estimator, whose MSE reaches this lower bound? More specifically, how can we obtain an estimator that is \textit{semiparametric efficient}, that is, an estimator which is asymptotically Gaussian and whose MSE satisfies the equality in \eqref{LB}?

Before answering this question, let’s start by reviewing the classic parametric case. It is well known that, under appropriate regularity conditions a parametric asymptotically efficient estimator is the Maximum Likelihood Estimator (MLE) \cite[Sect.\ 5.5]{vaart_1998}. For the symmetric location problem, the MLE for the center of symmetry $\theta_0$ in $\mathcal{P}_{\theta}$ in \eqref{par_mod} solves the likelihood equation:
\begin{equation}\label{est_eq}
	\left. \Delta_{g_0,n}(\theta) \right|_{\theta = \hat{\theta}_{n,ML}}= \left.\frac{1}{\sqrt{n}}l'(\theta)\right|_{\theta = \hat{\theta}_{n,ML}}= \left.\frac{1}{\sqrt{n}}\frac{d }{d \theta}\sum_{i=1}^{n}\ln g_0(x- \theta)\right|_{\theta = \hat{\theta}_{n,ML}} =0. 
\end{equation}
where $\Delta_{g_0,n}(\theta)$ is the central sequence previously defined in \eqref{cen_seq}. Moreover, we have that:
\begin{equation}
	\sqrt{n}(\hat{\theta}_{n,ML}-\theta_0) \overset{d}{\rightarrow} \mathcal{N}(0,I(\theta_0)^{-1}).
\end{equation}
Clearly, in the semiparametric context, the MLE cannot be implemented since $g_0$ is unknown. In particular, the semiparametric estimator of $\theta_0$ should be based on a \textit{$g_0$- and distribution-free} version of the central sequence $\Delta_{g_0,n}(\theta)$. The search for such a semiparametric version of the central sequence remains a problem that is still partly open for a general semiparametric model. However, for the simple univariate case of the symmetric location problem, we can make use of a groundbreaking result proposed in \cite{Hallin_Werker}. This result highlights the central importance of ordered statistics and their corresponding ranks and signs in semiparametric inference. So let’s take a brief detour to introduce these quantities and their properties in relation to the semiparametric symmetric location problem.

\subsection*{The ancillarity of the ranks and signs for the nuisance $g \in \mathcal{S}$}
The role of the ordered statistics can be understood through the derivation of the following two statistics for $g \in \mathcal{S}$ in the nonparametric sub-model $\mathcal{P}_{g} \subset \mathcal{P}_{\theta,g}$ in \eqref{nonpar_mod} \cite[Ch.\ 1]{Lehmann_TSH}:
\begin{itemize}
	\item A sufficient statistic for $g\in \mathcal{S}$ is a function of the observations $D_{\theta_0} = D_{\theta_0}(X_1,\ldots,X_n)$ that contains all the information about $g$ presents in the observations $\{X_i\}_{i=1}^n$. Formally, $D_{\theta_0}$ is sufficient for $g$ if the conditional joint density of $\{X_i\}_{i=1}^n$ given $D_{\theta_0}$ does not depend on $g$. This means that, once $D_{\theta_0}$ is known, the observations provide no additional information about $g$.
	\item An ancillary statistic $T_{\theta_0}=T_{\theta_0}(X_1,\ldots,X_n)$ of $g\in \mathcal{S}$ is a statistic whose distribution does not depend on $g$, that is $T_{\theta_0}$ provides no information about $g$ by itself.
\end{itemize}
Remarkably, sufficient and ancillary statistics are somehow complementary, as revealed by the Basu's theorems \cite{Basu_55,Basu_anc}. 

For the symmetric location problem, it can be shown that a (complete) sufficient and a (maximal) ancillary statistics for the functional parameter $g \in \mathcal{S}$ are respectively given by:
\begin{equation}\label{com_suf}
	D_{\theta_0} =\tonde{d_{(1)}(\theta_0),\ldots,d_{(n)}(\theta_0)},
\end{equation}
\begin{equation}\label{max_anc}
	T_{\theta_0} = \tonde{ r_{1}(\theta_0), \dots, r_{n}(\theta_0),  u_{1}(\theta_0), \dots, u_{n}(\theta_0)},
\end{equation}
where:
\begin{itemize}
	\item $d_{i}(\theta_0) \triangleq |X_i-\theta_0|$, $i = 1,\ldots,n$, and $d_{(1)}(\theta_0) <d_{(2)}(\theta_0)< \dots < d_{(n)}(\theta_0)$ are the ordered statistics of $d_{1}(\theta_0),\ldots,d_{n}(\theta_0)$ \cite[Chap. 13]{vaart_1998},
	\item $r_i(\theta_0) \triangleq \text{rank}(d_{i}(\theta_0))$ is the rank of $d_{i}$ in the set $\{d_{i}\}_{i=1}^n$ such that $d_{i} = d_{(r_{i})}$,
	\item $u_{i}(\theta_0) \triangleq \text{sign}(X_i-\theta_0)$ is the sign of $(X_i-\theta_0)$.
\end{itemize}
 Interested readers can find the proof of this result in the Section \virg{Proof of some technical results}.

\subsection*{A rank-based ($R$-) semiparametric estimator of $\theta_0$}
Let us now go back to our search of a semiparametric \textit{$g_0$- and distribution-free} version of the MLE. We saw in \eqref{est_eq} that the MLE can be obtained as the (supposed unique) root of the parametric central sequence $\Delta_{g_0,n}(\theta)$ in \eqref{cen_seq}. If we could find a \textit{$g_0$- and distribution-free} version of $\Delta_{g_0,n}(\theta)$, we may define the semiparametric counterpart of the MLE as the zero of such \textit{$g_0$- and distribution-free} central sequence. Since, as we have just discovered, $T_0$ in \eqref{max_anc} is an ancillary statistic for $g_0$ (then it does not contain any information about $g_0$ itself), intuitively such \textit{$g_0$- and distribution-free} central sequence should be a function of $T_0$ only and not of the observations $\{X_i\}_{i=1}^n$ themselves. Then, as proposed in \cite{Hallin_Werker}, we may take the conditional expectation of $\Delta_{g_0,n}(\theta)$ w.r.t. the ancillary statistic $T_0$. Such conditional expectation is the \virg{best prediction} of $\Delta_{g_0,n}(\theta)$ knowing the only $T_0$. To grasp its essence, below we break down this procedure into several steps for the specific case of the symmetric location problem:
\begin{enumerate}
	\item First of all, one can note that the function $\varphi_{g_0}(x- \theta)$ in \eqref{phi_fun} can be expressed as:
	\begin{equation}
		\varphi_{g_0}(x- \theta) = \mathrm{sign}(x- \theta) \varphi_{g_0}(|x- \theta|),\; \forall \theta \in \mathbb{R}.
	\end{equation}
	We can define an \textit{efficient central sequence} as:
	\begin{equation}
		\overline{\Delta}_{g_0,n}(\theta_0) \triangleq \frac{1}{\sqrt{n}}\sum_{i=1}^{n}\bar{s}_{\theta_0}(X_i),
	\end{equation} 
	where $\bar{s}_{\theta_0}$ is the efficient score defined in \eqref{def_eff_s_t}. As shown in \cite{Hallin_Werker}, $\overline{\Delta}_{g_0,n}(\theta_0)$ is \textit{asymptotically equivalent} (i.e.\ it is mean-square convergent) to the conditional expectation of the central sequence $\Delta_{g_0,n}(\theta_0)$ in \eqref{cen_seq} w.r.t.\ the ancillary statistic $T_{\theta_0}$:
	\begin{equation}\label{eff_cs}
		\begin{split}
			\overline{\Delta}_{g_0,n}(\theta_0) &= \evg{\Delta_{g_0,n}(\theta_0)|T_{\theta_0}} + o_{L_2}(1) \\
			& = \frac{1}{\sqrt{n}} \sum_{i=1}^n u_{i}(\theta_0) \e{\varphi_{g_0}(|X_i- \theta_0|)|r_1(\theta_0),\ldots,r_n(\theta_0)} + o_{L_2}(1)\\
			&  = \frac{1}{\sqrt{n}} \sum_{i=1}^n u_{i}(\theta_0) \eg{ \varphi_{g_0}\big(G_{0,+}^{-1}(U_{(r_i(\theta_0))})\big)} + o_{L_2}(1)\\
			& = \frac{1}{\sqrt{n}} \sum_{i=1}^n u_{i}(\theta_0) \eg{ K_{g_0}(U_{(r_i(\theta_0))})\big)} + o_{L_2}(1),
		\end{split}
	\end{equation}
	where $U_{(1)}<\cdots<U_{(n)}$ are the ordered statistics of $n$ random variables uniformly distributed on $(0,1)$ and, for notational simplicity, we introduced the function
	\begin{equation}
		K_{g_0}(q) \triangleq \varphi_{g_0}\big(G_{0,+}^{-1}(q)\big) = (\varphi_{g_0}\circ G_{0,+}^{-1})(q),\; q\in (0,1),
	\end{equation}
	where $G_{0,+}^{-1}$ is the \textit{quantile function}, i.e.\ the inverse of the cdf $G_{0,+}$ of $|X-\theta_0|$, such that $X-\theta_0$ is a random variable of distribution $G_0(x) = \int_{-\infty}^{x} g_0(\alpha)d\alpha$ with $g_0 \in \mathcal{S}$. Note that:
	\begin{equation}
		G_{0,+}(|x- \theta_0|) = \mathrm{Pr}\graffe{|X-\theta_0| \le |x- \theta_0|} = 2 G_0 (|x- \theta_0|) - 1,
	\end{equation}
	and then
	\begin{equation}
		G_{0,+}^{-1}(q) =  G_0^{-1}\left( \frac{1+q}{2} \right)\; q\in (0,1).
	\end{equation}
	Moreover, we have that
	\begin{equation}
		\overline{\Delta}_{g_0,n}(\theta_0) \overset{d}{\rightarrow} \mathcal{N}\big(0,\bar{I}(\theta_0|g_0)\big).
	\end{equation}
	
	\item Then \cite[Theo. 13.18]{vaart_1998} allows us to derive a \textit{rank-based} version of $\overline{\Delta}_{g_0,n}(\theta_0)$ as:
	\begin{equation}\label{app_cs}
		\widetilde{\Delta}_{g_0,n}(\theta_0) = \frac{1}{\sqrt{n}} \sum_{i=1}^n u_{i}(\theta_0) K_{g_0}\tonde{\frac{r_i(\theta_0)}{n+1}},
	\end{equation}
	that is asymptotically equivalent to $\overline{\Delta}_{g_0,n}(\theta_0)$, i.e.\ $\widetilde{\Delta}_{g_0,n}(\theta_0) = \overline{\Delta}_{g_0,n}(\theta_0) + o_{L_2}(1)$, and consequently is asymptotically normal with mean zero and variance $\bar{I}(\theta_0|g_0) =I(\theta_0)$:
		\begin{equation}
			\widetilde{\Delta}_{g_0,n}(\theta_0) \overset{d}{\rightarrow} \mathcal{N}(0,\bar{I}(\theta_0|g_0)).
		\end{equation}
	
	\item It must be noted that the expressions in \eqref{app_cs} still depend on the true and unknown density $g_0$ through $K_{g_0}$. To obtain a $g_0$-free, rank-based ($R$-) estimator, we need to get rid of this dependence. To this end, let us introduce the set $\mathcal{K}_\mathcal{S}$ of the \textit{rank score function} \cite[eq. (13.4)]{vaart_1998} to which $K_{g_0}$ belongs:
	\begin{equation}\label{K_score}
		\mathcal{K}_\mathcal{S} \triangleq \graffe{K_f : (0,1) \mapsto \mathbb{R}| K_f = \varphi_f \circ F_+^{-1}, f \in \mathcal{S}}.
	\end{equation}
	where $F_+^{-1}$ is the quantile function of a random variable $|X|$ such that $X \sim f \in \mathcal{S}$. We can then introduce a \textit{$g_0$- and distribution-free} version of the semiparametric efficient central sequence in \eqref{eff_cs}:
	\begin{equation}
		\begin{split}
			\overline{\Delta}_{f,n}(\theta_0) = \frac{1}{\sqrt{n}} \sum_{i=1}^n u_{i}(\theta_0) \eg{ K_f\big(U_{(r_i(\theta_0))}\big)},
		\end{split}
	\end{equation}
	and its \textit{rank-based} version:
	\begin{equation}\label{app_cs_f}
		\widetilde{\Delta}_{f,n}(\theta_0) = \frac{1}{\sqrt{n}} \sum_{i=1}^n  u_{i}(\theta_0)K_f\tonde{\frac{r_i(\theta_0)}{n+1}},\; f\in \mathcal{S}.
	\end{equation}
	
	From a direct application of \cite[Theo. 13.18]{vaart_1998}, we have that $\widetilde{\Delta}_{f,n}(\theta_0)$ is asymptotically equivalent to $\overline{\Delta}_{f,n}(\theta_0)$, that is $\widetilde{\Delta}_{f,n}(\theta_0) = \overline{\Delta}_{f,n}(\theta_0) + o_{L_2}(1)$. Let us introduce the term:
		\begin{equation}
			\nu(f,l) \triangleq  \int_{0}^{1}K_f(\alpha)K_{l}(\alpha)d\alpha,\; f,l \in \mathcal{S}.
		\end{equation}
		Then, again from \cite[Theo. 13.18]{vaart_1998}, $\widetilde{\Delta}_{f,n}(\theta_0)$ is asymptotically normal with mean zero and variance $\nu(f,f) \triangleq \int_0^1 K_f^2(\alpha) \, d\alpha$
		\begin{equation}
			\widetilde{\Delta}_{f,n}(\theta_0) \overset{d}{\rightarrow} \mathcal{N}(0,\nu(f,f)).
		\end{equation}
\end{enumerate}

This results allow us to propose a semiparametric counterpart of the MLE. In fact, in analogy with its \virg{parametric} definition given in \eqref{est_eq}, we may derive a rank-based $R$-estimator as the solution of:
\begin{equation}\label{r_lik_eq}
	\boxed{\left. \widetilde{\Delta}_{f,n}(\theta) \right|_{\theta = \hat{\theta}_{n,R}} = \left. \frac{1}{\sqrt{n}} \sum_{i=1}^n  u_{i}(\theta)K_f\tonde{\frac{r_{i}(\theta)}{n+1}}  \right|_{\theta = \hat{\theta}_{n,R}} = 0.}
\end{equation} 
The expression \eqref{r_lik_eq} shows that the related $\hat{\theta}_{n,R}$ is a function of the ranks and of the signs, then it is \textit{$g_0$- and distribution-free}. In fact, as shown for example in \cite[Lemma 13.17]{vaart_1998}, the rank vector $(r_{1}(\theta),\ldots,r_{n}(\theta))$ is uniformly distributed on the set of all $n!$ permutations of $\{1,\ldots,n\}$, while the sign vector $(u_{1}(\theta),\ldots,u_{n}(\theta))$ is uniformly distributed over $\{-1,1\}^n$ and then invariant with respect to $g_0$. 

We have finally reached the end of our journey. There is just one last step left: to show an explicit form of the estimator $\hat{\theta}_{n,R}$. It should be noted that solving the nonlinear equation \eqref{r_lik_eq} numerically is not an easy task, given that it involves ranks and signs that are not easy to handle from the perspective of numerical optimization. For the same reason, an asymptotic characterization of $\hat{\theta}_{n,R}$ is a prohibitive task. Fortunately, the concept of \virg{one-step} estimators offers us a solution, that however, comes with a price to pay: the distribution-freeness will be partially lost, as we will see in the next and final section.

\subsection*{One-step (OS) $R$-estimator}
The one-step procedure to derive efficient estimators is another contribution of Lucien Le Cam \cite{LeCam60}, \cite[Sect. 5.7]{vaart_1998}. In its most general form, the theory is quite abstract but if we impose stronger regularity conditions than those used by Le Cam, the intuition behind one-step estimation becomes straightforward. As we saw in the previous section, the estimator we are looking for is the one that solves the rank-based likelihood equation \eqref{r_lik_eq}. We therefore define the \textit{one-step (OS) $R$-estimator} $\hat{\theta}_{n,OS}$ as the root of \eqref{r_lik_eq} obtained by applying a single step of a Newton-Raphson-like method to $\widetilde{\Delta}_{f,n}(\theta)$. As initial guess, or in this context as \textit{preliminary} estimator, one may take any $g_0$-free, $\sqrt{n}$-consistent estimator \footnote{Note that, under Assumptions A1 and A2, such $\sqrt{n}$-consistent estimator always exists and can be obtained following, among others, the procedure described in \cite[Sect. 2.5]{BKRW}} of the parameter of interest.

Let us provide an operative definition of $\hat{\theta}_{n,OS}$ for our symmetric location problem. Let $\theta_n^\star$ be a $\sqrt{n}$-consistent preliminary estimator of $\theta_0$. For the problem at hand, we may take the sample median:
\begin{equation}
	\theta_n^\star = \mathrm{median}(X_1, \dots, X_n).
\end{equation}

Then an OS $R$-estimator can be expressed as \cite[Theo. 5.45]{vaart_1998}:
\begin{equation}\label{R_os}
	\boxed{\hat{\theta}_{n,OS} =  \theta_n^\star + \frac{1}{\sqrt{n}\widehat{\Psi}_{f,n}} \widetilde{\Delta}_{f,n}(\theta_n^\star),}
\end{equation} 
where $\widehat{\Psi}_{f,n} = \Psi_{f}(\theta_0) + o_{P_0}(1)$ is a consistent estimator of the quantity $\Psi_{f}(\theta_0)$ implicitly defined as: 
\begin{equation}\label{sort_der}
	\left| \widetilde{\Delta}_{f,n}(\theta_0 + n^{-1/2}h) - \widetilde{\Delta}_{f,n}(\theta_0) + \Psi_{f}(\theta_0)h \right|  = o_{P_0}(1),\; h \in \mathbb{R}.
\end{equation}
Intuitively, returning to the parallel with the Newton-Raphson method, $\Psi_{f}(\theta_0)$ represents \textit{a \virg{generalized derivative}} (but not the derivative in the classical sense that in general does not exist) at $\theta_0$ of our \virg{objective function} $\widetilde{\Delta}_{f,n}(\theta)$. Remarkably, $\Psi_{f}(\theta_0)$ can be explicitly expressed as \cite[Addendum 5.49]{vaart_1998}:
\begin{equation}
	\Psi_{f}(\theta_0) = \evg{\varphi_f(X)\bar{s}_{\theta_0}(X)},
\end{equation}   
where $\bar{s}_{\theta_0}$ is the efficient score defined in \eqref{eff_s_t}, that for the symmetric location problem is equal to the parametric score $s_{\theta_0}$ in \eqref{s_t}. Then, building upon the notation previously introduced, for the symmetric location problem, $\Psi_{f}(\theta_0)$ can be calculated (through a change of variable) as:
\begin{equation}\label{nu_ex}
	\Psi_{f}(\theta_0) = \int_{-\infty}^{+\infty} \varphi_f(x)\varphi_{g_0}(x)dG_0(x) = \nu(f,g_0).
\end{equation}
The expression \eqref{nu_ex} highlights an important fact: $\Psi_{f}(\theta_0)$ depends on $g_0$. Consequently, even though $\widetilde{\Delta}_{f,n}(\theta)$ in \eqref{app_cs_f} is distribution-free, the fact that $\Psi_{f}(\theta_0)$ is not implies that the OS estimator $\hat{\theta}_{n,OS}$ is only partially distribution-free.
	
Let us now focus on the search of the consistent estimator $\widehat{\Psi}_{f,n}$ for $\Psi_{f}(\theta_0)=\nu(f,g_0)$. For the symmetric location problem, a rank-based, consistent estimator $\widehat{\Psi}_{f,\alpha,n}$ is provided in \cite{Mckean}. A formal and in-depth description of this estimator and of its asymptotic properties is outside the scope of this Lecture Note. Here, we will simply highlight one of its features that may prove useful in its application. Specifically, $\widehat{\Psi}_{f,\alpha,n}$ depends on a hyperparameter $\alpha$ that allows us to adjust the trade-off between robustness and statistical efficiency. A numerical study, proposed in \cite{Mckean}, suggests $\alpha =0.1$ as default parameter. In general, if the data are heavy-tailed or contaminated by outliers, smaller $\alpha$ (e.g., 0.05) may improve robustness, while for clean or light-tailed data, larger $\alpha$ (e.g., 0.25) can enhance efficiency.

Finally, from a direct application of the Slutsky's lemma \cite[Lemma 2.8]{vaart_1998}, we have that:
\begin{equation}\label{R_con_dist}
	\boxed{\sqrt{n}(\hat{\theta}_{n,OS} - \theta_0)\overset{d}{\rightarrow} \mathcal{N}(0,\nu(f,g_0)^{-2}\nu(f,f))}.
\end{equation}
This result shows that the estimator obtained does not guarantee equality in \eqref{LB}, in other words it does not achieve the CRB for any $f \in \mathcal{S}$. To obtain a truly efficient estimator, one may consider estimating the nuisance function $g_0$ \textit{non-parametrically} and then using this estimator $\hat{g}_n$ to get the rank score function $K_{\hat{g}_n}$. In fact, if $\hat{g}_n = g_0 + o_{P_0}(1)$, then, following the previous derivation, we have that $\nu(\hat{g}_n,g_0) = \nu(g_0,g_0) + o_{P_0}(1)$ and this will ensure the efficiency of $\hat{\theta}_{n,OS}$ since:
\begin{equation}\label{R_eff}
	\sqrt{n}(\hat{\theta}_{n,OS} - \theta_0)\overset{d}{\rightarrow} \mathcal{N}(0,\nu(g_0,g_0)^{-1}),
\end{equation}  
where, by definition, $\nu(g_0,g_0)=\bar{I}(\theta_0|g_0) = I(\theta_0) = \evg{\varphi_{g_0}^2(X- \theta_0)}$. 

One might even consider using $\hat{g}_n$ to obtain a non-parametric MLE of $\theta_0$ and avoid the whole semiparametric machinery. Although this may seem like a good idea, this approach is hampered by the fact that, in general, to find a \virg{good} non-parametric estimator of $g_0$ may be an involved and cumbersome task (see e.g. \cite[Sect. 25.8]{vaart_1998}). For this reason, the semiparametric approach makes perfect sense: it allows us to obtain a $\sqrt{n}$-convergent estimator of the finite-dimensional parameter of interest without having to resort to a slow, unstable, computationally-demeaning and of questionable effectiveness non-parametric estimation of the nuisance function $g_0$.

Here we are at the final result: 
\begin{tcolorbox}[colback=gray!10!white,colframe=blue!50!black,arc=2mm,left=3pt,right=3pt,top=3pt,bottom=3pt]
	\textit{$\hat{\theta}_{n,OS}$ in \eqref{R_os} is a $g_0$-free and partially distribution-free estimator $\hat{\theta}_{n,OS}$ of the center of symmetry $\theta_0$ in the semiparametric model $\mathcal{P}_{\theta_0,g}$ in \eqref{sem_par} with the following properties:
			\begin{itemize}
				\item It is $\sqrt{n}$-consistent: $\sqrt{n}(\hat{\theta}_{n,OS} - \theta_0) = O_{P_0}(1)$ \footnote{The consistency is directly implied by the estimator's convergence in distribution in \eqref{R_con_dist}}
				\item It is asymptotically Gaussian with an asymptotic error covariance given by $\nu(f,g_0)^{-2}\nu(f,f)$,
				\item It is parametrically efficient (then equivalent to the MLE) for $\mathcal{S} \ni f = g_0$, that is, its asymptotic error covariance equate the parametric FI in \eqref{FI}: $\nu(g_0,g_0)=\bar{I}(\theta_0|g_0) = I(\theta_0)$.   
			\end{itemize}
	}
\end{tcolorbox}
\RV{Here a simple pseudo-code that describe the implementation of the proposed one-step $R$-estimator: \footnote{Code: \texttt{https://github.com/StefanoFor/R-estimator-for-the-symmetric-location-problem}} }
\begin{algorithm}
	\caption{One-step (OS) $R$-estimator}
	\begin{algorithmic}[1]
		\Require i.i.d.\ data $\{X_i\}_{i=1}^n$, choice of $f \in \mathcal{S}$, hyperparameter $\alpha$
		\Ensure $\hat{\theta}_{n,OS}$
		\State Preliminary estimation of $\theta_0$: $\theta_n^\star = \mathrm{median}(X_1, \dots, X_n)$
		\State Calculation of the ranks ans signs in \eqref{max_anc}: $\{r_{i}(\theta_n^\star)\}_{i=1}^n$ and $\{u_{i}(\theta_n^\star)\}_{i=1}^n$
		\State Derivation of the rank-score function $K_f \in \mathcal{K}_\mathcal{S}$ in \eqref{K_score}
		\State Calculation of $\widetilde{\Delta}_{f,n}(\theta_n^\star) = \frac{1}{\sqrt{n}} \sum_{i=1}^n  u_{i}(\theta_n^\star)K_f\tonde{\frac{r_i(\theta_n^\star)}{n+1}}$ in \eqref{app_cs_f}
		\State Implementation of $\widehat{\Psi}_{f,\alpha,n} \equiv \widehat{\Psi}_{f,\alpha,n}(\{X_i\}_{i=1}^n,\theta_n^\star)$ in \cite{Mckean}
		\State \Return $\hat{\theta}_{n,OS} =  \theta_n^\star + \frac{1}{\sqrt{n}\widehat{\Psi}_{f,\alpha,n}} \widetilde{\Delta}_{f,n}(\theta_n^\star)$ in \eqref{R_os}
	\end{algorithmic}
\end{algorithm}

Before moving on to numerical simulations, let us provide a classic example of rank score function $K_f \in \mathcal{K}_\mathcal{S}$: the Gaussian-based rank score function $K_\mathcal{N}$. Specifically, let's take $f = \mathcal{N}(0,1)$. Then the function $\varphi_f$ in \eqref{phi_fun} is given by $\varphi_\mathcal{N}(x) = x$. Moreover, by defining as $\Phi^{-1}$ the quantile function of the Gaussian distribution, we have:
\begin{equation}\label{K_G}
	F_{+,\mathcal{N}}^{-1}(q) =  \Phi^{-1}\left( \frac{1+q}{2} \right)\; q\in (0,1).
\end{equation}
Consequently, Gaussian-based rank score function is given by: $K_\mathcal{N}(q) = F_{+,\mathcal{N}}^{-1}(q)$ for $ q\in (0,1)$. Note that, according to \eqref{R_eff}, the choice of the Gaussian-based rank score function $K_\mathcal{N}$ ensures that, if the data are Gaussian distributed, $\hat{\theta}_{n,OS}$ will not suffer any loss of efficiency compared with the standard Gaussian-based MLE.

\section*{Numerical study}
To convince readers that what has just been presented is not only of theoretical value, but can also be easily implemented and used directly in applications, in this section, we report a simulative study of the presented OS $R$-estimator for the center of symmetry $\theta_0$.

We generate $n=100$ i.i.d. observations, centered at $\theta_0 = 6$, from the following two densities:
\begin{itemize}
	\item \textit{Case 1}: a Student-$t$ density with $\nu$ degrees of freedom, \RV{i.e.\ $g_0(x) = p_{t_\nu}(x) =
	\frac{\Gamma\left(\frac{\nu+1}{2}\right)}
	{\sqrt{\nu\pi}\Gamma\left(\frac{\nu}{2}\right)}
	\tonde{
		1+\frac{x^2}{\nu}}^{-\frac{\nu+1}{2}}.$ \cite[Sect. 5.2]{Chap_back_JP}}. 
	We recall that when $\nu$ is close to $0$, the $t$-distributed data are heavily non-Gaussian. On the other hand, when $\nu \rightarrow \infty$, Student-$t$ distribution converges to the Gaussian one.
	\item \textit{Case 2}: a \textit{Generalised Gaussian} (GG) density with shape parameter $s>0$ and scale $b>0$, i.e.\ \RV{$g_0(x) = p_{GG_{s,b}}(x)= 
	\frac{s}
	{\sqrt{2}\,b^{\frac{1}{2s}}
		\Gamma(1/2s)}
	\exp\left[
	-\frac{x^{2s}}{2^sb}
	\right],$ \cite[Sect. 5.3]{Chap_back_JP}}. Note that, for $0<s<1$ the tails of the GG distribution are heavier w.r.t.\ the Gaussian one, for $s=1$ we have the Gaussian distribution, and for $s > 1$ the tails of the GG distribution are lighter w.r.t.\ the Gaussian one. In our simulations, we set $b=1$.  
\end{itemize}

 In the two above-mentioned cases, we compare the lower bound in \eqref{LB} with the estimated MSE, that is $\widehat{\text{MSE}}(\theta_0) =\frac{1}{\text{MC}}\sum_{m=1}^{\text{MC}}(\hat{\theta}_i-\theta_0)^2$, of four estimators of $\theta_0$ over $\text{MC}=10^5$ Monte Carlo iterations:
\begin{itemize}
	\item The sample mean $\hat{\theta}_{n,Mean} = \frac{1}{n}{\sum_{i=1}^n X_i}$: \textit{$g_0$-free}, but \textbf{not distribution-free} (it is in fact a function of the observations $\{X_i\}_{i=1}^n$) \textbf{neither robust to large deviations from the center of symmetry $\theta_0$}.
	\item The sample median $\hat{\theta}_{n,Med} = \mathrm{median}(X_1, \dots, X_n)$ and the Huber $M$-estimator $\hat{\theta}_{n,H}$ \cite{Huber_M}: \textit{$g_0$-free and robust to large deviations from $\theta_0$}, but \textbf{not distribution-free}. For the interested reader, we recall that Huber $M$-estimator can be obtained as the solution to the following equation:
	\begin{equation}
	\left. \sum\nolimits_{i=1}^n
		\psi_c(X_i-\theta)\right|_{\theta = \hat{\theta}_{n,H}} 
		=
		0,\; \text{where}\; \psi_c(v)
		=
		\left\lbrace \begin{array}{ll}
			v,
			& |v|\leq c,\\
			c\,\text{sign}(v),
			& |v|>c.
		\end{array}\right., 
	\end{equation}
	and $c$ is a hyperparameter controlling the trade-off between efficiency (at Gaussianity) and robustness to outliers. In particular, for $c\rightarrow \infty$, $\hat{\theta}_{n,H}$ tends to the sample mean, while for $c\rightarrow 0$ it tends to the sample median. In our simulations, we choose $c=0.9$. Note that both $\hat{\theta}_{n,Med}$ and $\hat{\theta}_{n,H}$ are functions of the observations $\{X_i\}_{i=1}^n$, so they are not distribution-free.     
	\item The OS $R$-estimator $\hat{\theta}_{n,OS}$ in \eqref{R_os} with the Gaussian rank score function $K_\mathcal{N}$: \textit{$g_0$-free, partially distribution-free and robust to large deviations from $\theta_0$}\footnote{The robustness large deviations of $\hat{\theta}_{n,OS}$ is guaranteed by the robustness to outliers of the chosen preliminary estimator $\theta_n^\star$.},
\end{itemize}
Note that the Fisher Information $I(\theta_0)$ in \eqref{FI} can be explicitly evaluated as:	
\begin{itemize}
	\item for the Student-$t$ distribution: $I(\theta_0) = \frac{\nu + 1}{\nu+3}$,
	\item for the GG distribution: $I(\theta_0) = 4s^2 (b2^s)^{-1/s}\frac{\Gamma( 2 - 1/(2s) )}{\Gamma(1/(2s))}$.
\end{itemize}

\begin{figure}[htbp]
	\centering
	\includegraphics[width=0.4\textwidth]{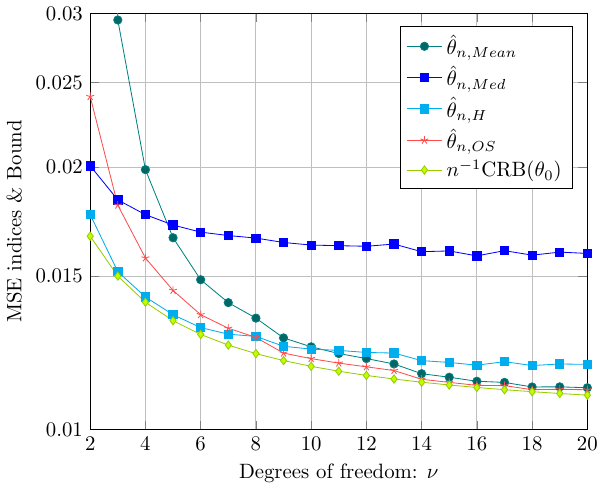}
	\caption{Case 1: $t$-distributed observations.}\label{fig1}
\end{figure}

\begin{figure}[htbp]
	\centering
	\includegraphics[width=0.4\textwidth]{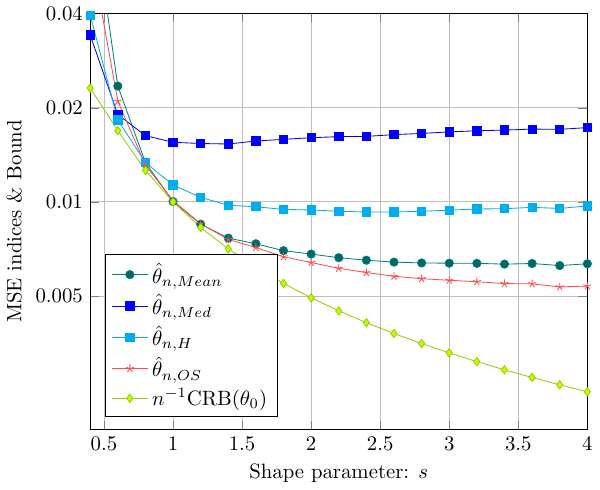}
	\caption{Case 2: $GG$-distributed observations ($b=1$).}\label{fig2}
\end{figure}

Here some considerations and take-away messages about the simulation results.
\begin{itemize}
	\item \textit{Robustness to large deviation from $\theta_0$}: As might be expected, the MSE of $\hat{\theta}_{n,Mean}$ explodes when the data are highly non-Gaussian, i.e. for small values of $\nu$ for the Case 1 (Fig. \ref{fig1}) and for $0<s<1$ in Case 2 (Fig. \ref{fig2}). On the contrary, $\hat{\theta}_{n,Med}$, $\hat{\theta}_{n,H}$ and $\hat{\theta}_{n,OS}$ have an MSE that remains bounded for every value of $\nu$.
	\item \textit{Statistical efficiency}: We can note the good performance of $\hat{\theta}_{n,H}$ which remains close to the $\text{CRB}(\theta_0)$, particularly for highly non-Gaussian data. However, for the Case 1, the $\hat{\theta}_{n,OS}$ performs better for values of $\nu > 7$. A similar behavior is observed in Case 2 where $\hat{\theta}_{n,OS}$ outperforms $\hat{\theta}_{n,H}$ for $s > 0.8$.
	\item \textit{Statistical efficiency at Gaussian data}: As predicted by \eqref{R_eff}, the numerical analysis has confirmed that $\hat{\theta}_{n,OS}$ achieves the $\text{CRB}(\theta_0)$ when the data are Gaussian. This is obtained when $\nu \rightarrow \infty$ for the Case 1 and when $s=1$ for the Case 2.  
	\item \textit{Distributional robustness}: It is worth recalling that the true \virg{added value} of $\hat{\theta}_{n,OS}$ compared with $\hat{\theta}_{n,Mean}$, $\hat{\theta}_{n,Med}$ and $\hat{\theta}_{n,H}$ is in the fact that it is distribution-free (at least in the form given in \eqref{r_lik_eq}), since as we have seen, it is a function of the ranks and the signs that have a distribution independent of the actual $g_0$. As we have demonstrated analytically in this Lecture Note and as can be seen from the simulations, despite this invariance with respect to $g_0$, $\hat{\theta}_{n,OS}$ is able to exhibit good statistical efficiency. The study of this surprising efficiency property of ranks and of the signs is an active field of research in non-parametric statistics.
\end{itemize}

\section*{What we have learned}
The aim of this Lecture Note is to provide an introduction to semiparametric statistics to the SP community as a framework that reconciles statistical efficiency and distributional robustness. By estimating finite-dimensional parameters (e.g., the center of symmetry) while treating infinite-dimensional nuisances (e.g., noise density) as unknown, we may obtain semiparametric estimators that are at the same time (almost) distribution-free and (almost) efficient. Taking the symmetric location problem as an example, we have seen how this apparent magic is made possible by the existence of an ancillary statistic for the infinite-dimensional nuisance parameter, which in our case is represented by the ranks and signs of the observations.

We hope this Lecture Note has highlighted the significant impact that semiparametric statistics can have on SP applications. We also hope that many other researchers in this community will join the effort to advance semiparametric statistics from both a theoretical and an applied perspective.
%
%
%
%
%

\section*{Proof of some technical results}
\subsection*{Sufficiency and completeness of $D_{\theta_0}$ for $g \in \mathcal{S}$ in $\mathcal{P}_{g}$}
The joint density of the ordered observations $X_{(1)},\ldots,X_{(n)}$ is (see e.g.\ \cite[Ex.\ 2.4.1]{Lehmann_TSH}):
\begin{equation}
	f(x_{(1)},\ldots,{x_{(n)}};g)=n!\mathds{1}_{\{x_1<x_2<\cdots<x_n\}}\prod_{i=1}^n g(x_i-\theta_0) = n!\mathds{1}_{\{x_1<x_2<\cdots<x_n\}}\prod_{i=1}^n g(|x_i-\theta_0|).
\end{equation}
where the second equality follows from the symmetry of $g$. Consequently, we can write
\begin{equation}
	f(x_{(1)},\ldots,{x_{(n)}};g)=h(x_1,\dots,x_n) \prod_{i=1}^n g(d_{i}(\theta_0)),
\end{equation}
where $h(x_1,\dots,x_n)= n!\mathds{1}_{\{x_1<x_2<\cdots<x_n\}}$. Then, by the Factorization Theorem, $D_{\theta_0}$ is sufficient for $g$ in $\mathcal{P}_{g}$ in \eqref{nonpar_mod}.

Let us now move to the completeness. The statistic $D_{\theta_0}$ is said to be \textit{complete} for $g$ in $\mathcal{P}_{g}$ if, for every bounded measurable function $L$:
\begin{equation}\label{complet}
	E_g\{L(D_{\theta_0})\} = 0, \; \forall g\in \mathcal{S} \Rightarrow L(D_{\theta_0})=0,\; \mathcal{P}_{g}-\text{a.s.}\;\forall g\in \mathcal{S}.
\end{equation}
We start by noticing that, from the fact that $X_1,\ldots,X_n$ are i.i.d.\ observations and from the definition of $D_{\theta_0}$ given in \eqref{com_suf}, the term $E_g\{L(D_{\theta_0})\}$ can be expressed as:
\begin{equation}\label{con_com}
	E_g\{L(D_{\theta_0})\} = \int_{\mathbb{R}^n} L(|x_{(1)}-\theta_0|,\ldots,|x_{(n)}-\theta_0|)\prod_{i=1}^n  g(x_i - \theta_0)\,dx_1\ldots dx_n = 0
\end{equation}
for all symmetric densities $g \in \mathcal{S}$. Let $B \in \mathcal{B}((0,+\infty))$ be a Borel set of positive measure, i.e. $\lambda(B)>0$. Let $g_B(x) \triangleq \frac{\mathds{1}_B + \mathds{1}_{-B}}{2\lambda(B)}$ (where $-B \triangleq \{-b, b\in B\}$) be a symmetric density. Then, by substituting such $g_B(x)$ in \eqref{con_com}, we get $\int_{B^n} L(|x_{n(1)}-\theta_0|,\ldots,|x_{n(n)}-\theta_0|)\,dx_1\ldots dx_n = 0$. Since $B$ is arbitrary, then $L(D_{\theta_0})=0$ $\lambda$-a.e.\ To conclude we note that, by assumption, any measure $G$ (associated to a density $g$) is absolutely continuous with respect to the Lebesgue measure $\lambda$. Consequently, the fact that $L(D_{\theta_0})=0$ $\lambda$-a.e.\ implies $L(D_{\theta_0})=0$ $G$-a.e.\ and then $L(D_{\theta_0})=0$, $\mathcal{P}_{g}$-a.s.\ for all $g\in\mathcal{S}$. Then, according to \eqref{complet}, we can conclude that $D_{\theta_0}$ is complete.

\subsection*{Maximal ancillarity of $T_{\theta_0}$ in $\mathcal{P}_{g}$}
From the properties of the ordered statistics and the relevant rank and sign statistics (see e.g. \cite[Ch. 13]{vaart_1998}), it follows that the statistics $T_{\theta_0}$ in \eqref{max_anc} and the sufficient and complete statistic $D_{\theta_0}$ in \eqref{com_suf} are independent. Then, as a consequence of the so-called Basu's first theorem \cite[Theo. 1]{Basu_55}, we have that $T_{\theta_0}$ is an ancillary statistic for $g$ in $\mathcal{P}_{g}$ in \eqref{nonpar_mod}. Moreover, by noticing that, for the symmetric location problem, each observation can be decomposed as:
\begin{equation}
	X_i = \theta_0 + d_{(r_i(\theta_0))}u_{i}(\theta_0),
\end{equation}  
the following relation among $\sigma$-algebras generated by the observations $X_1,\ldots,X_n$ and the statistics $D_{\theta_0}$ and $T_{\theta_0}$ holds true:
\begin{equation}\label{sigma_S_T}
	\sigma(X_1,\ldots,X_n) = \sigma(\sigma(D_{\theta_0}) \cup \sigma(T_{\theta_0})).
\end{equation} 
Thanks to \eqref{sigma_S_T}, and since, as shown before, $D_{\theta_0}$ in \eqref{com_suf} is sufficient and complete for $g$ and $T_{\theta_0}$ in \eqref{max_anc} is ancillary for $g$ in $\mathcal{P}_{g}$, we can directly apply the so-called Basu's third theorem \cite[Theo. 7]{Basu_anc} to conclude that $T_{\theta_0}$ in \eqref{max_anc} is a maximal ancillary.

\bibliographystyle{IEEEtran}
\bibliography{ref_semipar_eff_estim}

\begin{thebibliography}{10}
\providecommand{\url}[1]{#1}
\csname url@samestyle\endcsname
\providecommand{\newblock}{\relax}
\providecommand{\bibinfo}[2]{#2}
\providecommand{\BIBentrySTDinterwordspacing}{\spaceskip=0pt\relax}
\providecommand{\BIBentryALTinterwordstretchfactor}{4}
\providecommand{\BIBentryALTinterwordspacing}{\spaceskip=\fontdimen2\font plus
\BIBentryALTinterwordstretchfactor\fontdimen3\font minus
  \fontdimen4\font\relax}
\providecommand{\BIBforeignlanguage}[2]{{%
\expandafter\ifx\csname l@#1\endcsname\relax
\typeout{** WARNING: IEEEtran.bst: No hyphenation pattern has been}%
\typeout{** loaded for the language `#1'. Using the pattern for}%
\typeout{** the default language instead.}%
\else
\language=\csname l@#1\endcsname
\fi
#2}}
\providecommand{\BIBdecl}{\relax}
\BIBdecl

\bibitem{For_Delmas_Ollila_TIT}
S.~Fortunati, J.-P. Delmas, and E.~Ollila, ``Nuisance parameters and
  elliptically symmetric distributions: A geometric approach to parametric and
  semiparametric efficiency,'' \emph{IEEE Transactions on Information Theory},
  vol.~72, no.~9, pp. 6902--6927, 2026.

\bibitem{LeCam60}
L.~{Le {C}am}, ``Locally asymptotically normal families of distributions,'' in
  \emph{Univ. California Publ. Statist.}, vol.~3, 1960, pp. 37--98.

\bibitem{vaart_1998}
A.~W. {van der {V}aart}, \emph{Asymptotic Statistics}, ser. Cambridge Series in
  Statistical and Probabilistic Mathematics.\hskip 1em plus 0.5em minus
  0.4em\relax Cambridge University Press, 1998.

\bibitem{BKRW}
P.~Bickel, C.~Klaassen, Y.~Ritov, and J.~Wellner, \emph{Efficient and Adaptive
  Estimation for Semiparametric Models}.\hskip 1em plus 0.5em minus 0.4em\relax
  Johns Hopkins University Press, 1993.

\bibitem{Chap_sem}
\BIBentryALTinterwordspacing
S.~Fortunati, ``Semiparametric estimation in elliptical distributions,'' in
  \emph{Elliptically Symmetric Distributions in Signal Processing and Machine
  Learning}, J.-P. Delmas, M.~N. El~Korso, S.~Fortunati, and F.~Pascal,
  Eds.\hskip 1em plus 0.5em minus 0.4em\relax Cham: Springer Nature
  Switzerland, 2024, pp. 149--185. [Online]. Available:
  \url{https://hal.science/hal-04141679v3}
\BIBentrySTDinterwordspacing

\bibitem{Hallin_Werker}
M.~Hallin and B.~J.~M. Werker, ``Semi-parametric efficiency,
  distribution-freeness and invariance,'' \emph{Bernoulli}, vol.~9, no.~1, pp.
  137--165, 2003.

\bibitem{Lehmann_TSH}
E.~L. Lehmann and J.~P. Romano, \emph{Testing Statistical Hypotheses}, 3rd~ed.,
  2008.

\bibitem{Basu_55}
D.~Basu, ``On statistics independent of a complete sufficient statistic,''
  \emph{Sankhyā: The Indian Journal of Statistics (1933-1960)}, vol.~15,
  no.~4, pp. 377--380, 1955.

\bibitem{Basu_anc}
------, ``The family of ancillary statistics,'' \emph{Sankhyā: The Indian
  Journal of Statistics (1933-1960)}, vol.~21, no. 3/4, pp. 247--256, 1959.

\bibitem{Mckean}
J.~W. McKean and T.~P. Hettmansperger, ``A robust analysis of the general
  linear model based on one step ${R}$-estimates,'' \emph{Biometrika}, vol.~65,
  no.~3, pp. 571--579, 1978.

\bibitem{Chap_back_JP}
\BIBentryALTinterwordspacing
J.-P. Delmas, ``Background on real and complex elliptically symmetric
  distributions,'' in \emph{Elliptically Symmetric Distributions in Signal
  Processing and Machine Learning}, J.-P. Delmas, M.~N. El~Korso, S.~Fortunati,
  and F.~Pascal, Eds.\hskip 1em plus 0.5em minus 0.4em\relax Cham: Springer
  Nature Switzerland, 2024, pp. 1--34. [Online]. Available:
  \url{https://hal.science/hal-04217510v5}
\BIBentrySTDinterwordspacing

\bibitem{Huber_M}
P.~J. Huber, ``{Robust Estimation of a Location Parameter},'' \emph{The Annals
  of Mathematical Statistics}, vol.~35, no.~1, pp. 73 -- 101, 1964.

\end{thebibliography}

\vspace{-1cm}
%
%
%

\begin{IEEEbiographynophoto}{Stefano Fortunati}
(stefano.fortunati@telecom-sudparis.eu) received the PhD at the University of Pisa in 2012, where he stayed as researcher until 2019. He also spent one year as visiting researcher at the Signal Processing Group at Technische Universität Darmstadt. From 2019 to 2024, he was with the Laboratoire des Signaux et Systèmes (L2S) at Université Paris-Saclay. From 2024 he is associate professor at Télécom SudParis/SAMOVAR. He is the co-recipient of the IEEE AESS 2021 Barry Carlton Award and of the 2019 EURASIP JASP Best Paper Award. Dr. Fortunati’s professional expertise encompasses different areas of statistics and the statistical signal processing, with particular focus on performance bounds, misspecification theory, robust and semiparametric statistics.
\end{IEEEbiographynophoto}

\vfill

\end{document}